\documentclass{article}

\usepackage{arxiv}

\usepackage[utf8]{inputenc} 
\usepackage[T1]{fontenc}    
\usepackage{hyperref}       
\usepackage{url}            
\usepackage{booktabs}       
\usepackage{amsfonts}       
\usepackage{nicefrac}       
\usepackage{microtype}      
\usepackage{lipsum}		
\usepackage{graphicx}
\usepackage{natbib}
\usepackage{doi}
\usepackage{subcaption}

\title{Nonparametric Estimation of Conditional Survival Function with Time-Varying Covariates Using DeepONet 
\thanks{This work was supported in part by NIH grants RF1 AG075107 and UCI Alzheimer's Disease Research Center Grant P30 AG066519, and by NSF grant DMS 2412746.}
}

\author{ 
{
Bingqing Hu}
\\
	Department of Statistics \\ University of California, Irvine \\ Irvine, CA 92697 \\
	\texttt{bingqih2@uci.edu} \\
	\And
    {
    Bin Nan} \\
	Department of Statistics \\ University of California, Irvine \\ Irvine, CA 92697 \\
	\texttt{nanb@uci.edu} \\
}

\renewcommand{\shorttitle}{Survival Estimation via DeepONet}

\begin{document}
\maketitle

\begin{abstract}
Traditional survival models often rely on restrictive assumptions such as proportional hazards or instantaneous effects of time-varying covariates on the hazard function, which limit their applicability in real-world settings. We consider the nonparametric estimation of the conditional survival function, which leverages the flexibility of neural networks to capture the complex, potentially long-term non-instantaneous effects of time-varying covariates. In this work, we use Deep Operator Networks (DeepONet), a deep learning architecture designed for operator learning, to model the arbitrary effects of both time-varying and time-invariant covariates. Specifically, our method relaxes commonly used assumptions in hazard regressions by modeling the conditional hazard function as an unknown nonlinear operator of entire histories of time-varying covariates. The estimation is based on a loss function constructed from the nonparametric full likelihood for censored survival data. Simulation studies demonstrate that our method performs well, whereas the Cox model yields biased results when the assumption of instantaneous time-varying covariate effects is violated. We further illustrate its utility with the ADNI data, for which it yields a lower integrated Brier score than the Cox model. 
\end{abstract}


\section{Introduction}

Survival prediction is of increasing importance, and estimating the conditional survival function from censored data nonparametrically is of great interest. Recent advances in machine learning and deep learning have overcome certain limitations of traditional semiparametric survival analysis methods \citep{cox, AFT}, such as Gradient Boosting Machines (GBM) \citep{survival_ensembles, boosting} and Random Survival Forests (RSF) \citep{rsf}, which are ensemble methods based on survival trees \citep{survival_tree}; and DeepSurv \citep{deepsurv} and DeepHit \citep{rnnsurv}, which are based on neural networks. Although some of them \citep{survival_ensembles, boosting, deepsurv} take more flexible non-linear forms, they still keep certain model structures in the framework of either the Cox model or the accelerated failure time model. Also, they estimate a risk score to summarize the features rather than estimating the conditional survival function directly. DeepHit divides the continuous time into a series of evenly partitioned time intervals and outputs the probability of an event occurring in each time interval. 
RSF is a flexible approach that uses the logarithmic rank test to split the data and directly output the survival probabilities at the observed time points. However, these methods are primarily developed for time-invariant covariates, although extensions are possible for handling time-varying covariates. 
In particular, \cite{tv_ml} found that the computational cost is extremely high for RSF when dealing with time-varying covariates. 

Following the same data expansion technique used to fit a Cox model, \cite{hu&nan} proposed a nonparametric approach using neural networks to directly estimate the conditional survival function given time-varying covariates. It takes time-varying covariates, baseline covariates, and a corresponding time interval of a partition formed by all observed time points as input to a feedforward neural network (FNN) and the logarithm of the conditional hazard function as output. The method uses a loss function determined by the nonparametric full likelihood, thus does not assume any specific model structure. 

To the best of our knowledge, all known methods for time-varying covariates, such as those discussed above, assume that the effects of covariates on the conditional hazard function are instantaneous. That is, the instantaneous hazard given entire covariate paths depends only on the values of covariates observed at the current time point. Such an assumption can be unrealistic, however, and considering flexible non-instantaneous effects of time-varying covariates on the hazard function can be of great interest and importance in addressing such issues, for example, when the hazard is related to an arbitrary cumulative exposure or certain delayed covariate effects. A nonparametric approach for non-instantaneous covariate effects provides the most flexible way of estimating how covariates influence survival, which captures complex temporal dynamics and eliminates model biases in survival prediction. 

In this work, we further generalize the method of \cite{hu&nan} to account for arbitrary effects of entire covariate histories of time-varying covariates. The time-varying covariates are functional inputs and the conditional survival function (or equivalently the conditional hazard function) becomes an operator. We propose applying the recently developed deep operator network, DeepONet \citep{deeponet}, to estimate the unknown operator.  
To our knowledge, the method is the first to model arbitrary effects of time-varying covariates in estimating the conditional survival function nonparametrically with the fewest assumptions.

\section{Methodology}

Consider survival times of $n$ subjects, which can be right-censored. We are interested in estimating the survival time distribution given a set of covariates, and some or all of them are time-varying covariates. For subject $i$, we denote the time-varying covariate vector as $X_i(t)$. We further denote the underlying failure time as $T_i$, and the underlying censoring time as $C_i$. We assume that the failure time possesses a Lebesgue density and that the censoring time is independent of failure time given covariates. 
Let the observed time be $Y_i = \min\{T_i, C_i\}$ and the failure indicator be $\Delta_i = I(T_i\le C_i)$. The observations $\{Y_i,\Delta_i,\widetilde X_i(\cdot): \ i=1,\dots,n\}$ are independent and identically distributed,  where $\widetilde X_i(t)$ denotes the covariate history up to time $t$, that is, $\widetilde X_i(t)=\{X_i(s), 0\le s \le t\}$. Let $f[t | \widetilde X_i(\infty)]$ and $f_C[t|\widetilde X_i(\infty)]$ be the conditional density functions of $T_i$ and $C_i$, respectively; and $S[t|\widetilde X_i(\infty)]$ and $S_C[t|\widetilde X_i(\infty)]$ be the conditional survival functions of $T_i$ and $C_i$, respectively.

We further assume that future covariate values do not influence the current risk. In other words, the risk of the event of interest occurring at time $t$ only depends on the current and past values of covariates, not on future values, so the covariates are external. Otherwise, the time-varying covariate is called the internal covariate \citep{Kalbfleisch-Prentice-2002}. Note that the conditional survival
probability is not well-defined if there is an internal covariate. Then the conditional hazard function of $T_i$ can be written as:
\begin{equation}
\label{eq: assump_history}
    \lambda\left[t \left | \widetilde X_i(\infty)\right.\right]=\lambda\left[t \left | \widetilde X_i(t)\right.\right].
\end{equation}

To meet the positivity constraint for the hazard function, we transform the hazard function with the logarithm to ease numerical implementations. Denote $$h[t, \widetilde X_i(t)] = \log \lambda[t|\widetilde X_i(t)].$$ Then the conditional cumulative hazard function given covariate history has the following form:
$$
   \Lambda\left[t \left | \widetilde X_i(\infty)\right.\right]= \Lambda\left[t \left | \widetilde X_i(t)\right.\right]=\int_{0}^t \lambda\left[s \left | \widetilde X_i(s)\right.\right]ds=\int_{0}^t e^{h[s,\widetilde X_i(s)]}ds,
$$
and the conditional survival function is given by
\begin{equation}
    \label{eq:S_4}
 S\left[t \left | \widetilde X_i(\infty) \right.\right]= \exp\left\{-\Lambda \left[t \left | \widetilde X_i(t)\right]\right.\right\} = \exp\left\{-\int_{0}^t e^{h[s,\widetilde X_i(s)]}ds \right\}.
\end{equation}

Equation (\ref{eq:S_4}) is always a valid survival function for any function $h[s,\widetilde X_i(s)]$. Once the function $h$ is estimated, we can easily determine the conditional survival function estimator.

\subsection{Data Structure and DeepONet}

Clearly $h[s,\widetilde X_i(s)]$ is an operator since $\widetilde X_i(s)$ can be viewed as a functional input. We propose to estimate  $h[s,\widetilde X_i(s)]$ nonparametrically using the DeepONet \citep{deeponet}. Following \cite{deeponet}, we partition the time axis evenly into $m$ intervals at $t_0\equiv 0 < t_1 < t_2< \cdots < t_m\equiv\tau$ and consider covariate values on these grid points, where  $\tau$ is a finite upper bound of the considered support of the survival time, often taken to be the largest observed failure time in a dataset in practice. 
Note that $h[s,\widetilde X_i(s)] = h[s,\widetilde X_i(\tau)]$ for external covariates $X_i(\cdot)$, we input only the historical values of the covariates up to the time point $s$ and mask future values with $0$. This turns out to be a valid approach to apply the DeepONet, which can be seen from the result of the universal approximation theorem for operators given in Equation (\ref{eq:UAT}). Then following \cite{hu&nan} we expand the observed covariates $x_i(\cdot)$ of each subject $i$ into $m$ rows of inputs, as shown in Table ~\ref{tab:data_structure_arbitrary}. 

\bigskip
\begin{center}
[Table ~\ref{tab:data_structure_arbitrary} is around here.]
\end{center}
\bigskip

\begin{figure}[htb!]
    \centering
    \includegraphics[width=0.5\textwidth]{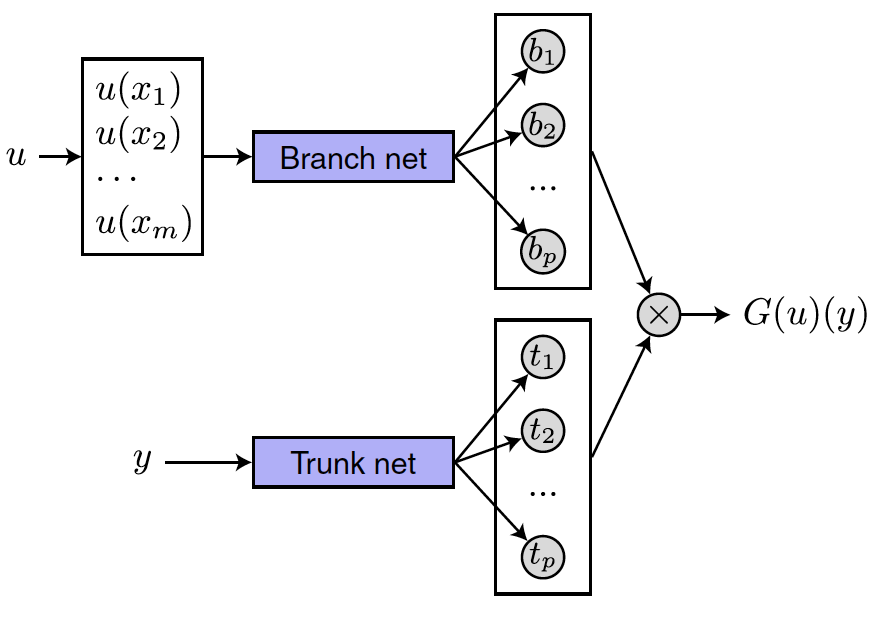}
    \caption{The unstacked DeepONet Structure. $u$, $y$, $G(u)(y)$ in this figure correspond to $x$, $s$, $h(x)(s)$, respectively.}
    \label{fig:deeponet}
\end{figure}

For implementing the DeepONet, we directly
employ a classical neural network structure -- either fully connected FNN or convolutional neural networks (CNN), and concatenate $\widetilde x(s)$ and $s$ together as the network input. Note the difference to the existing work that used such concatenatation \citep{biganzoli,rnnsurv,hu&nan,cbnn} only to the covariate value at the current time point, i.e., $[s,x_i(s)]$, whereas we consider the entire covariate history up to the current time point which naturally leads to the formulation of the conditional survival distribution using an operator. 
Recently, using neural networks to approximate operators has drawn increasing attention in the literature \citep{deeponet,fno_universal}.  \cite{deeponet} showed that the deep operator network can learn various explicit operators, such as integrals and fractional Laplacians, as well as implicit operators that represent deterministic and stochastic differential equations. Our problem can be seen as an operator learning problem. Using the operator notation, we denote $h[s,\widetilde x_i(s)]$ as  $h[\widetilde x_i(s)](s)$ and use both of them exchangeably in this article.  Specifically, we use an unstacked DeepONet structure to estimate $h[\widetilde x_i(s)](s)$. A DeepONet consists of two sub-networks --  the branch net for encoding the input function at $m$ sensors $[x(t_0), x(t_1), ..., x(t_{m-1})]^T$ and the trunk net for encoding the location $s$ for the output function \citep{deeponet}. See Figure~\ref{fig:deeponet} for an illustration where the input is denoted as $(u(\cdot), y)$ and the output is denoted as $G(u)(y)$.  

\bigskip
\begin{center}
[Figure~\ref{fig:deeponet} is around here.]
\end{center}
\bigskip

The architecture of branch and trunk neural networks is inspired by the universal approximation theorem for operators \citep{operator_universal}. The theorem states that two fully connected neural networks with a single hidden layer, combined by a vector dot product of the outputs, are able to approximate any continuous nonlinear operator with arbitrary accuracy. Specifically, suppose that \(\sigma\) is a continuous non-polynomial function, \(X\) is a Banach space, \(K_1 \subset X\) and \(K_2 \subset {R}^d\) are compact sets in \(X\) and \({R}^d\), respectively, \(V\) is a compact set in \(C(K_1)\), and \(G\) is a nonlinear continuous operator which maps \(V\) into \(C(K_2)\). Here $C(K)$ is the Banach space of all continuous
functions defined on $K$ equipped with the uniform norm. For any \(\epsilon > 0\), there are positive integers \(n\), \(p\), and \(m\), constants \(c_i^k\), \(\xi_{ij}^k\), \(\theta_i^k\), \(w_k \in {R}\), \(x_j \in K_1\), \(i = 1, \ldots, n\), \(k = 1, \ldots, p\), and \(j = 1, \ldots, m\), such that
\begin{equation} \label{eq:UAT}
\left| G(u)(y) - \sum_{k=1}^p \sum_{i=1}^n c_i^k \sigma \left( \sum_{j=1}^m \xi_{ij}^k u(x_j) + \theta_i^k \right) \sigma (w_k \cdot y + \zeta_k) \right| < \epsilon
\end{equation}
holds for all \(u \in V\) and \(y \in K_2\).

A generalized version of the approximation theorem (Theorem 2 in \cite{deeponet}) states that for any nonlinear continuous operator $G$ and any $\epsilon>0$, there exist positive integers $m$, $p$, continuous vector functions $g: {R}^m \rightarrow {R}^p$, $f: {R}^d \rightarrow {R}^p$, and $x_1, x_2, \ldots, x_m$, such that, 
\begin{equation}
    |G(u)(y)-\langle g[u(x_1),u(x_2),...,u(x_m)],f(y)\rangle| < \epsilon
\end{equation}
holds for all $u$ and $y$, where $\langle \cdot, \cdot \rangle$ denotes the dot product in ${R}^p$. The functions $g$ and $f$ can be chosen as diverse classes of neural networks, for example, FNN or CNN.

\subsection{Full Likelihood and Discretized Loss}

We propose using the full likelihood to build the loss function. Given observations $\{y_i,\delta_i,x_i(\cdot)\}$, the full likelihood function is
\begin{eqnarray*}
    \label{eq:likelihood_4}
        L_n &=& \prod_{i=1}^n \left\{f[y_i | \widetilde{x}_i(y_i)]S_C[y_i | \widetilde{x}_i(y_i)]\right\}^{\delta_i}\left\{f_C[y_i | \widetilde x_i(y_i)]S[y_i | \widetilde{x}_i(y_i)]\right\}^{1-\delta_i} \nonumber\\
        &\propto& \prod_{i=1}^n \lambda[y_i |\widetilde{x}_i(y_i)]^{\delta_i}S[y_i |\widetilde{x}_i(y_i)] \nonumber \\ 
        &=& \prod_{i=1}^n \exp\{h[\widetilde{x}_i(y_i)](y_i)\delta_i\}\exp\left\{-\int_0^{y_i} e^{h[\widetilde{x}_i(t)](t)}dt \right\}.
\end{eqnarray*}
The log-likelihood is given by 
\begin{equation} \label{eq:log-likelihood_4}
        \ell_n = \sum_{i=1}^n \left\{h[\widetilde{x}_i(y_i)](y_i)\delta_i-\int_0^{y_i} e^{h[\widetilde x_i(t)](t)}dt\right\}.
\end{equation}
By calculating the above integrals numerically using the summation on a partition of $[0, \tau]$ with $m$ subintervals, we obtain the loss function that is the discretized negative log-likelihood, scaled by the sample size:
\begin{eqnarray}
\label{eq:loss_deeponet}
        loss(h) &=& \frac{1}{n} \sum_{i=1}^n \sum_{j=1}^m I(t_{j-1} \le y_i) \Big\{ e^{h[\widetilde{x}_{i}(t_{j-1})](t_{j-1})}(t_j - t_{j-1})   \\ 
        && \qquad - \, h[\widetilde{x}_i(t_{j-1})](t_{j-1}) \delta_{ij}\Big\}, \nonumber
\end{eqnarray}
where $\delta_{ij} = I(t_j > y_i)\delta_i$. Note that $\delta_{ij}$ is always 0 until the event happens.

\subsection{Details in Network Architectural and Hyperparameter Tuning}
\label{sec:nn_arch}
We implement the DeepONet using the python package DeepXDE \citep{deepxde}. The package is primarily for solving ordinary and partial differential equations. It supports five tensor libraries as backends. To adapt the method to our problem, we modify the code in DeepXDE using its TensorFlow 2.x backend, including using our customized loss function and modifying the branch net structure for the CNN method. We train the neural networks on a system equipped with an AMD Ryzen 9 8945HS processor with Radeon 780M Graphics at 4.00 GHz and 32GB of RAM.

For the trunk net, we use the simple one layer FNN structure because there is only one scalar input of time. For the branch net, we use either FNN or CNN to summarize the time-varying covariates, where the FNN has two fully connected hidden layers (a.k.a. dense layer), and the CNN has two blocks of 1D convolutional layers \citep{1dcnn} and max pooling layers followed by one fully connected layer.  When using the FNN structure for the branch net, we concatenate the time-invariant covariates to the time-varying covariates directly; when using the CNN structure, however, it is not reasonable to concatenate two types of covariates so we first input the time-varying covariates into CNN and then concatenate the time-invariant covariates with the CNN output before going through the final fully connected layer. 
The masked future values of $0$ for the time-varying covariates (see Table \ref{tab:data_structure_arbitrary}) will not contribute to the output of the FNN branch net because the multiplication with those future time input-weights will be $0$, regardless of the weight values (see Equation (\ref{eq:UAT})). To ensure that future covariate values do not contribute to the output of the CNN branch net, we use casual padding \citep{causal_padding} that automatically neglects the future time points. Our simulations given in the next section show that the CNN branch net performs better, where we consider one time-varying covariate and two time-invariant covariates. This is not surprising because CNN preserves the temporal neighborhood information of the time-varying covariates.

We use an independent validation dataset to decide when to stop during training and use another independent test dataset to tune the hyperparameters. Specifically, in each training, we stop when the validation loss no longer decreases, and we choose the hyperparameter combination with the smallest test loss. In our simulations, we choose the hyperparameters from the following combinations:

\begin{itemize}
    \item number of nodes in each dense layer: $[32,64,128,256]$
    \item number of filters in each Conv1D layer: $[16,32,64]$
    \item pool size: $[4,8]$
    \item learning rate: $[0.01,0.001,0.0001]$
    \item batch size: $[100,500,1000]$ 
\end{itemize}

\section{Simulations}

We generate data from a model that has cumulative covariate effects on the conditional hazard function. Thus the Cox model under the usual assumption $\lambda [t|\widetilde X_i(t)] = \lambda [t|X_i(t)]$ is misspecified in this setup.

First, we generate a time-varying covariate on a fine grid of $[0, 100]$. For $t \in \{0, \Delta s, 2\Delta s,....,100\}$ with $\Delta s = 0.1$, $i \in \{1,2,...n\}$, we generate random variables $\alpha_{i1},\dots,\alpha_{i5}$ independently from the $\rm{Uniform} \, (0,1)$ distribution and construct a time-varying covariate as the following:
\begin{eqnarray*}    
    X_{i}(t)&=&\alpha_{i1}+\alpha_{i2}\sin(2\pi t/\tau)+\alpha_{i3}\cos(2\pi t/\tau)+\alpha_{i4}\sin(4\pi t/\tau) \\
    && \qquad + \, \alpha_{i5}\cos(4\pi t/\tau).
 \end{eqnarray*}
The sample path $x_i(t)$ of the above covariate is a left-continuous step function with right limit on the grid. Then we generate two time-invariant covariates $Z_{i} \sim \rm{Bernoulli} \, (0.5)$ and $W_{i} \sim \rm{Normal} \, (0,1)$. 
For a set of covariate values, we determine the following conditional hazard function of the survival time:
    \begin{eqnarray*}
        \lambda(t  | \widetilde x_i(t), z_i, w_i) 
        &=& 0.05*\exp\Bigg(w_i+z_i+0.01*\sum_{s\le t} x_i(s)\Delta s \\
        && \qquad + \, 0.01*\sum_{s\le t} x_i^2(s) z_i\Delta s \Bigg),  
    \end{eqnarray*}
then evaluate the conditional cumulative hazard function and the conditional survival function on the grid: 
\begin{eqnarray}
        \Lambda(t |\widetilde x_i(t)) &=& \Delta s \sum_{s\le t}\lambda_i(s | x_i(s)), \label{eq:Lambda}\\
        S(t |\widetilde x_i(t)) &=& \exp\left\{-\Lambda_i(t | \widetilde x_i(t))\right\}. \label{eq:Survival}
\end{eqnarray}

To generate the failure time $T_i$, we first generate $U_i$ from the $\rm{Uniform}\, (0,1)$ distribution, then set $T_i=\sup\left\{t:S_i(t |\widetilde x_i(t)) \ge U_i\right\}$.
We generate the censoring time $C_i$ by taking the minimum of a randomly generated value from the $\rm{Exponential} \, (50)$ distribution and $99$ which yields around 20\% censoring rate. Then we have  $Y_i = T_i \wedge C_i$ and $\Delta_i=I(T_i\le C_i)$.

Note that such generated time variable takes discrete values on the fine grid, which can be viewed as a numerical approximation of the continuous time variable. But due to its discrete nature, the exact survival function should be expressed as a product integral determined by the above hazard function (\ref{eq:Lambda}). We find, however, the above survival function given in (\ref{eq:Survival}) approximates the exact survival function adequately in our simulation studies. Using the product integral in (\ref{eq:Survival}) provides a way to analyze discrete survival times.

We independently generate training sets and validation sets each with a sample size of $n=2000$, then use the time-varying covariate values taken on $m=250$ equally spaced grid points on $[0,100]$ to train the DeepONet. Here, $m$ is treated as another hyperparameter taking values from the set $\{ 100, 200, 300, 400, 500\}$ and $m=250$ is the median of the optimal $m$ values obtained from 10 independent datasets. The hyperparameters of the neural networks chosen from a dataset with a sample size of $n=2000$ are as follows: the number of nodes in each dense layer (except for the last layer in branch and trunk net) is 128, the number of filters in each Conv1D layer is 16, the pool size is 8 with the same stride, and the learning rate is 0.001 and batch size is 1000. We then fix these hyperparameter values during the simulation runs.  ``Relu" function is used as the activation function between the hidden layers, and linear function is used for the final output so that the output value is not constrained. ``Adam" is used as the optimizer. In the last layer before doing the dot product, we use $p=10$ nodes as suggested in \cite{deeponet}.
Finally, we use the fitted model to estimate the conditional survival curves given 9 different sets of newly generated covariates. We repeat the process $N=200$ times, plot the sample average and 90\% confidence bands of the estimated conditional survival curves on these 9 sets of covariates. 

\begin{figure}[htb!]
\centering
\begin{subfigure}{0.3\textwidth}
\includegraphics[width=0.9\linewidth, height=3cm]{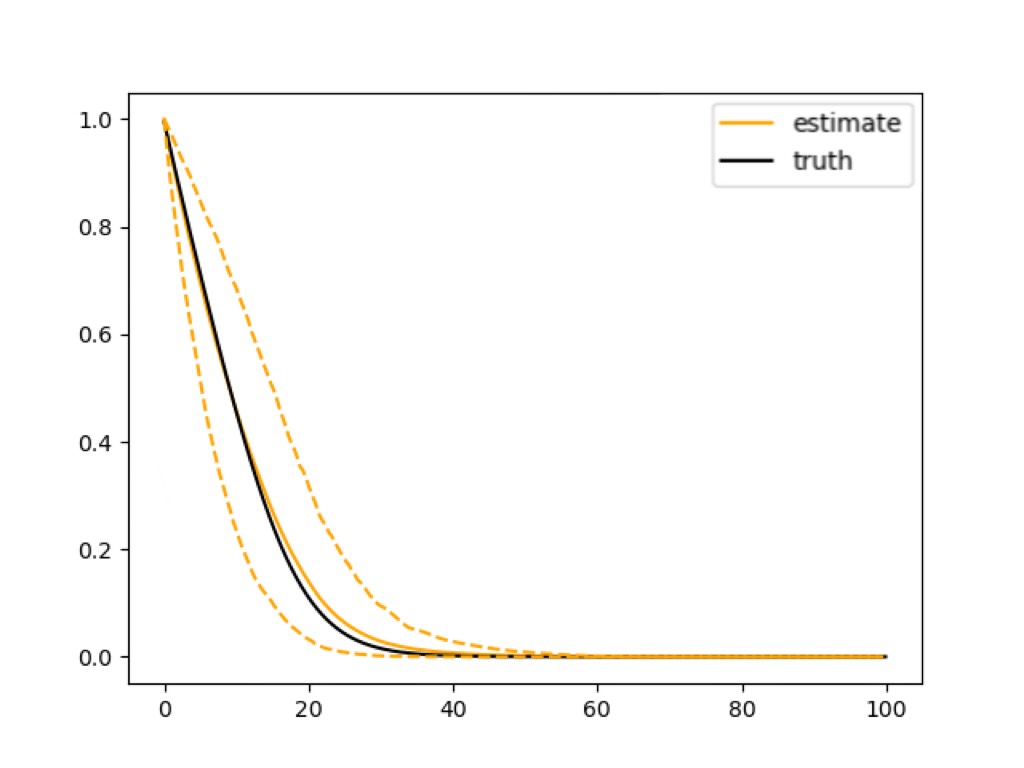} 
\end{subfigure}\hfil
\begin{subfigure}{0.3\textwidth}
\includegraphics[width=0.9\linewidth, height=3cm]{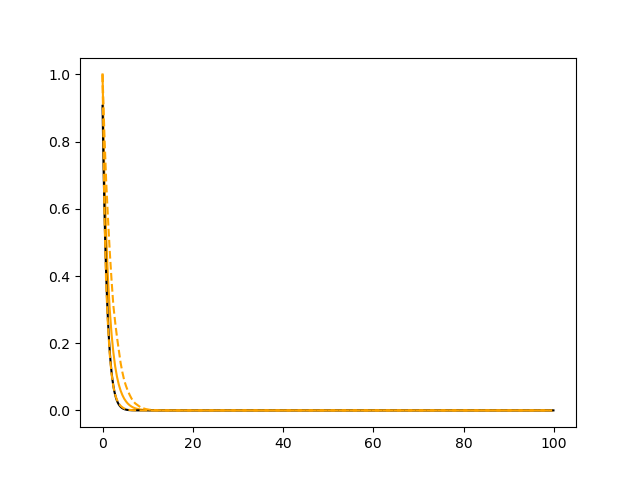}
\end{subfigure}\hfil
\begin{subfigure}{0.3\textwidth}
\includegraphics[width=0.9\linewidth, height=3cm]{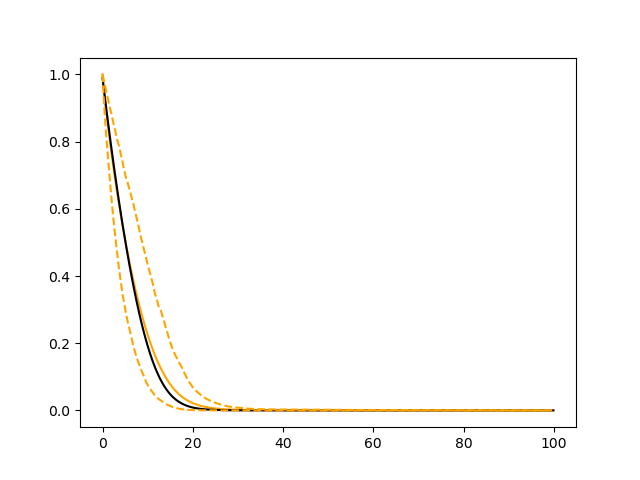}
\end{subfigure}\hfil
\begin{subfigure}{0.3\textwidth}
\includegraphics[width=0.9\linewidth, height=3cm]{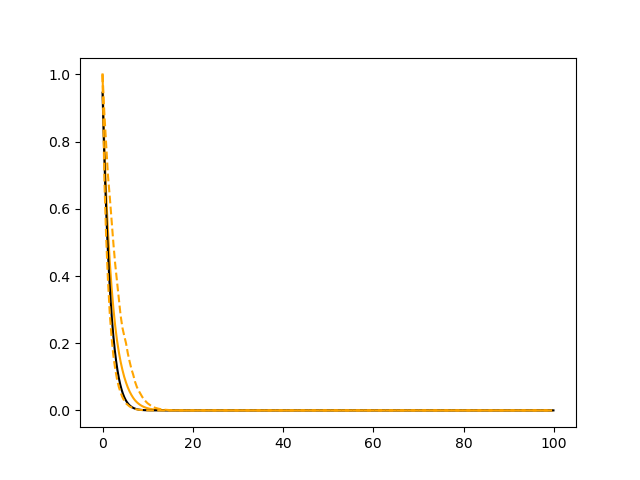}
\end{subfigure}\hfil
\begin{subfigure}{0.3\textwidth}
\includegraphics[width=0.9\linewidth, height=3cm]{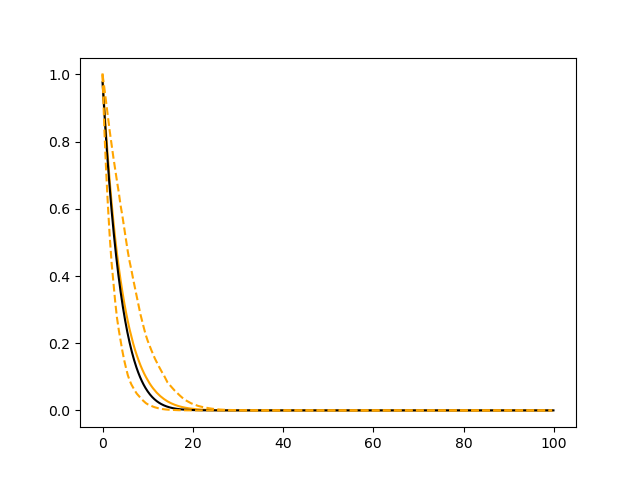} 
\end{subfigure}\hfil
\begin{subfigure}{0.3\textwidth}
\includegraphics[width=0.9\linewidth, height=3cm]{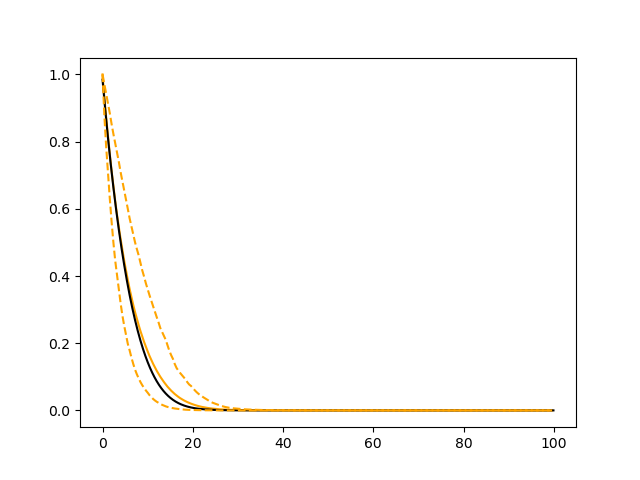} 
\end{subfigure}\hfil
\begin{subfigure}{0.3\textwidth}
\includegraphics[width=0.9\linewidth, height=3cm]{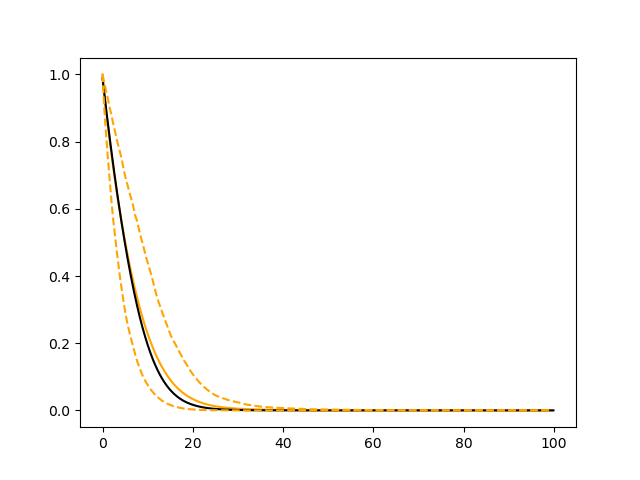} 
\end{subfigure}\hfil
\begin{subfigure}{0.3\textwidth}
\includegraphics[width=0.9\linewidth, height=3cm]{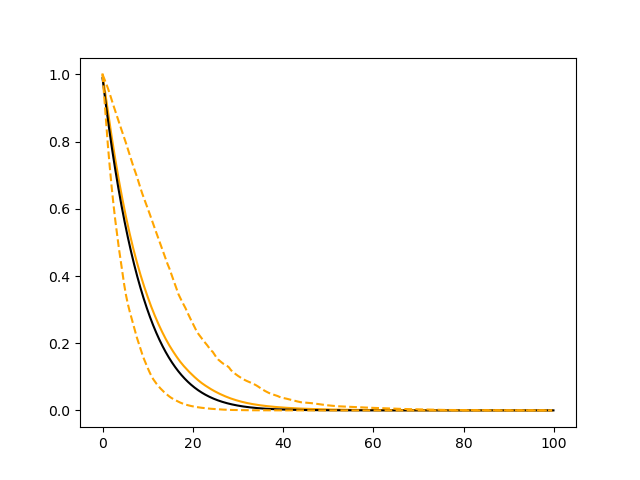} 
\end{subfigure}\hfil
\begin{subfigure}{0.3\textwidth}
\includegraphics[width=0.9\linewidth, height=3cm]{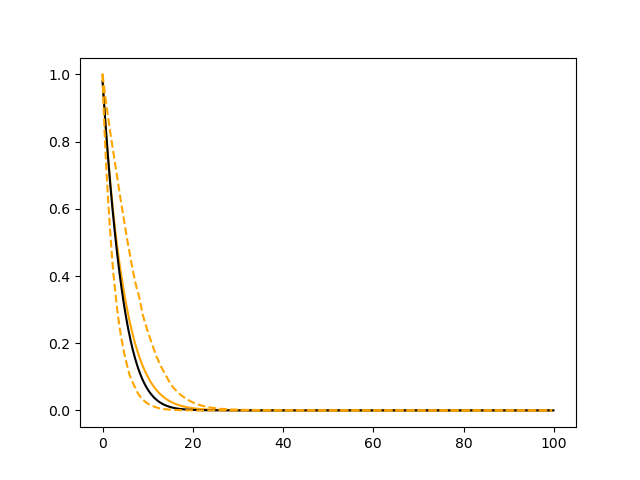} 
\end{subfigure}\hfil
\caption{Conditional survival curves for 9 different sets of covariates estimated by the DeepONet with FNN branch net and sample size $n = 2000$.}
\label{fig:arbitrary}
\end{figure}

\begin{figure}[htb!]
\centering
\begin{subfigure}{0.3\textwidth}
\includegraphics[width=0.9\linewidth, height=3cm]{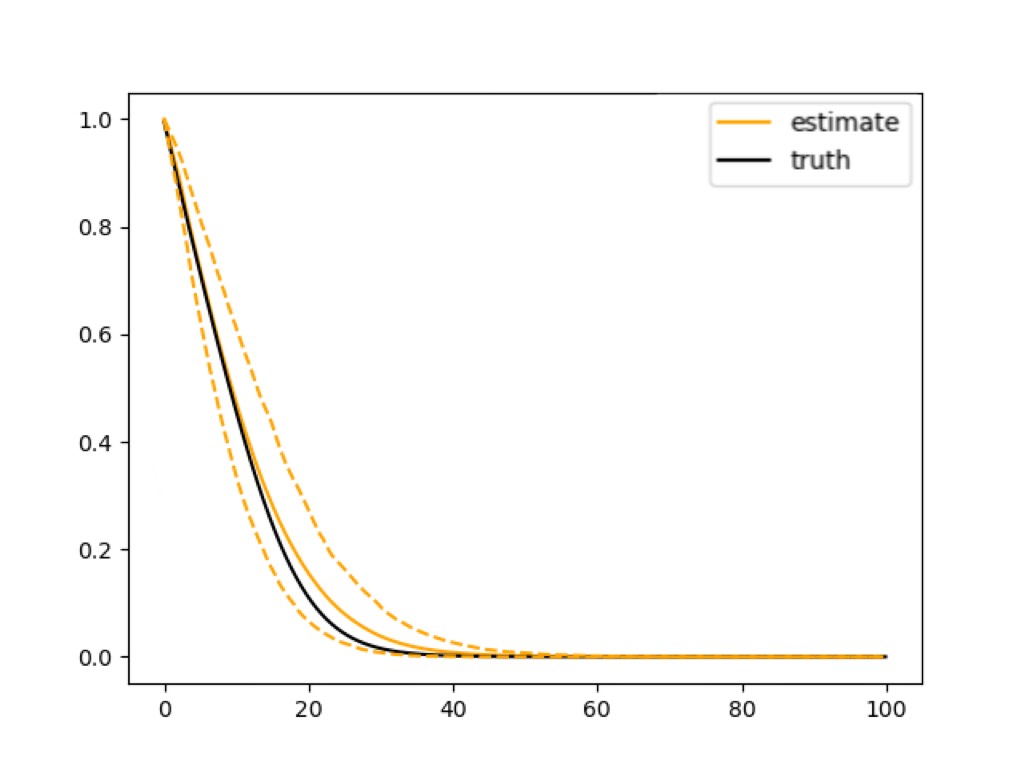} 
\end{subfigure}\hfil
\begin{subfigure}{0.3\textwidth}
\includegraphics[width=0.9\linewidth, height=3cm]{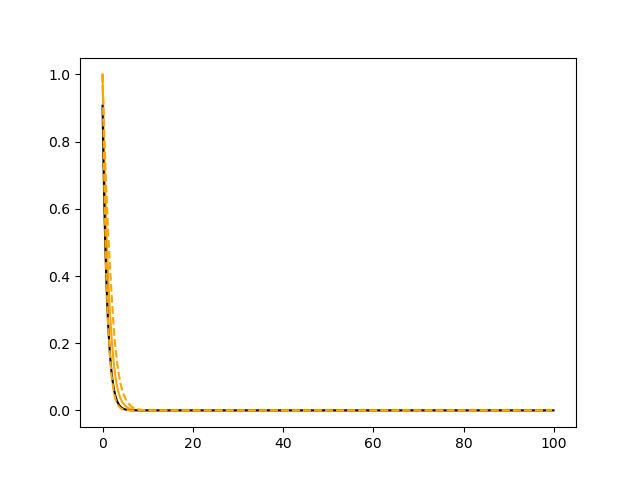}
\end{subfigure}\hfil
\begin{subfigure}{0.3\textwidth}
\includegraphics[width=0.9\linewidth, height=3cm]{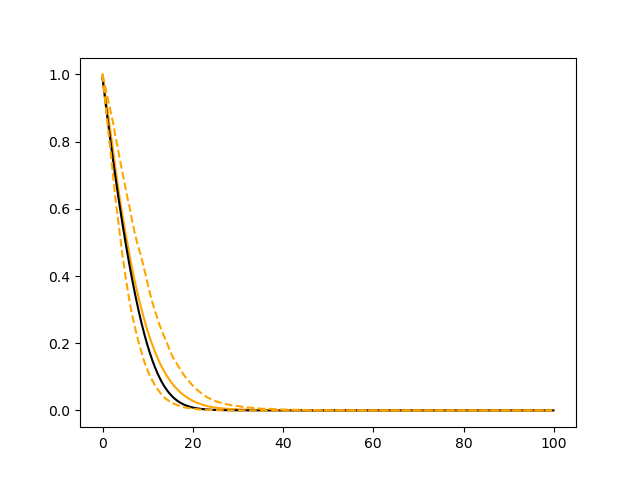}
\end{subfigure}\hfil
\begin{subfigure}{0.3\textwidth}
\includegraphics[width=0.9\linewidth, height=3cm]{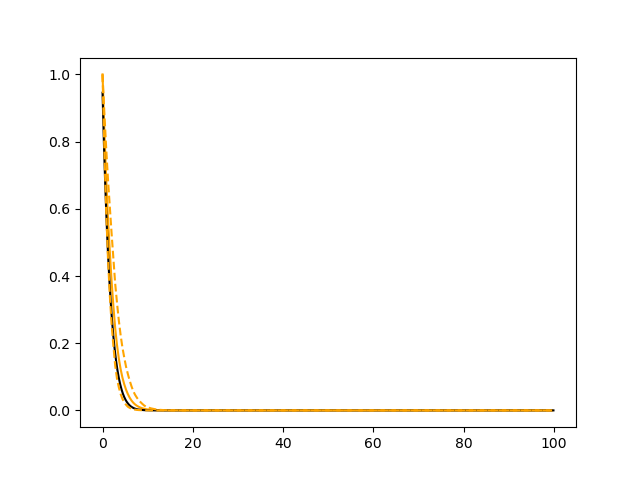}
\end{subfigure}\hfil
\begin{subfigure}{0.3\textwidth}
\includegraphics[width=0.9\linewidth, height=3cm]{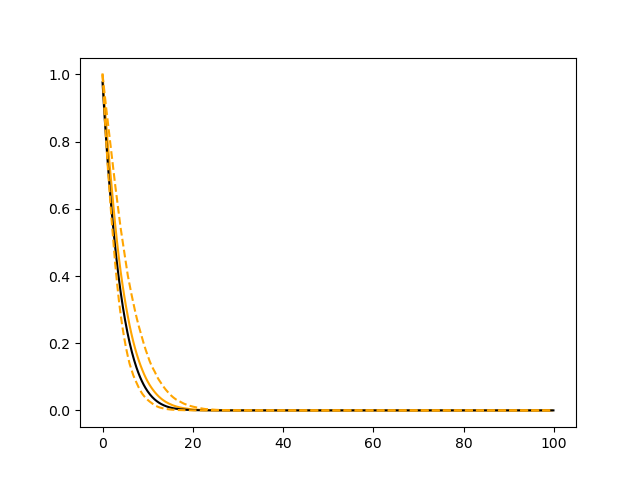} 
\end{subfigure}\hfil
\begin{subfigure}{0.3\textwidth}
\includegraphics[width=0.9\linewidth, height=3cm]{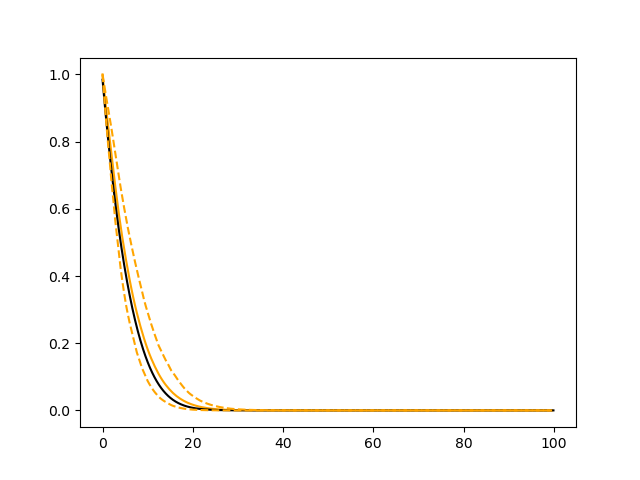} 
\end{subfigure}\hfil
\begin{subfigure}{0.3\textwidth}
\includegraphics[width=0.9\linewidth, height=3cm]{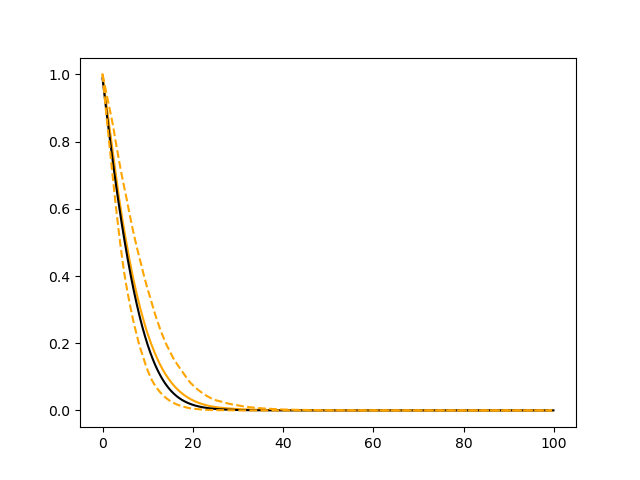} 
\end{subfigure}\hfil
\begin{subfigure}{0.3\textwidth}
\includegraphics[width=0.9\linewidth, height=3cm]{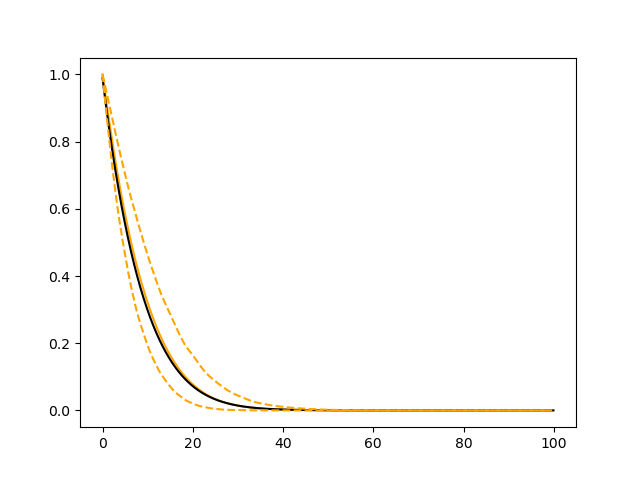} 
\end{subfigure}\hfil
\begin{subfigure}{0.3\textwidth}
\includegraphics[width=0.9\linewidth, height=3cm]{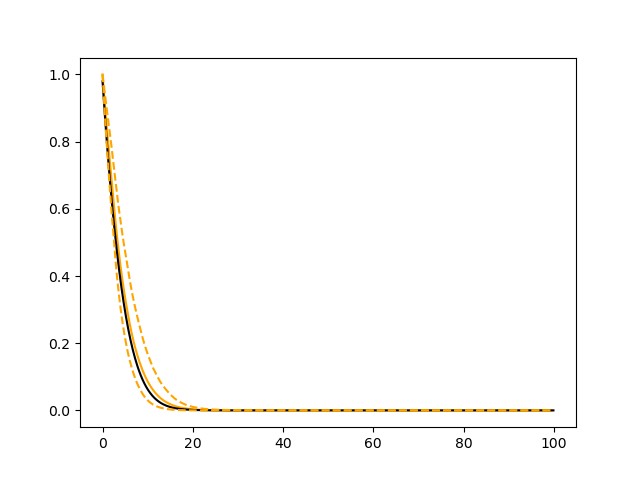} 
\end{subfigure}\hfil
\caption{Conditional survival curves for 9 different sets of covariates estimated by the DeepONet with CNN branch net and sample size $n=2000$. }
\label{fig:arbitrary_cnn}
\end{figure}

\begin{figure}[htb!]
\centering
\begin{subfigure}{0.3\textwidth}
\includegraphics[width=0.9\linewidth, height=3cm]{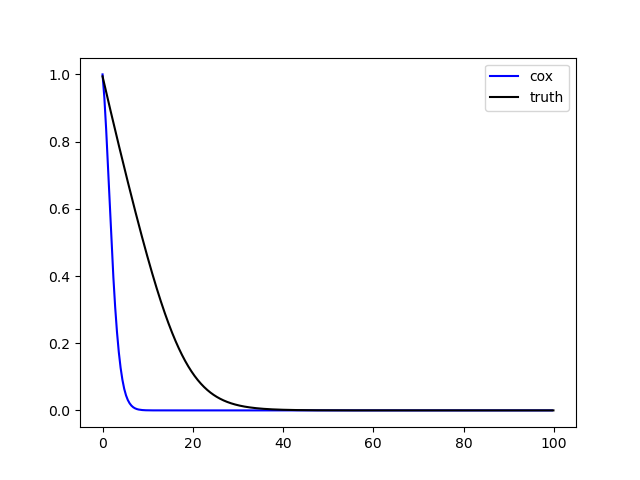} 
\end{subfigure}\hfil
\begin{subfigure}{0.3\textwidth}
\includegraphics[width=0.9\linewidth, height=3cm]{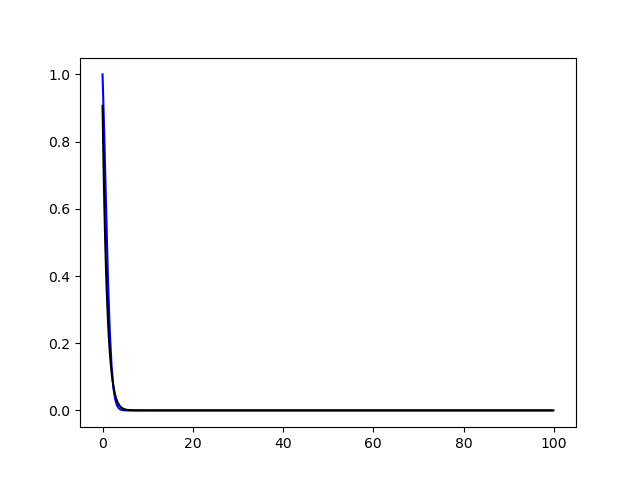}
\end{subfigure}\hfil
\begin{subfigure}{0.3\textwidth}
\includegraphics[width=0.9\linewidth, height=3cm]{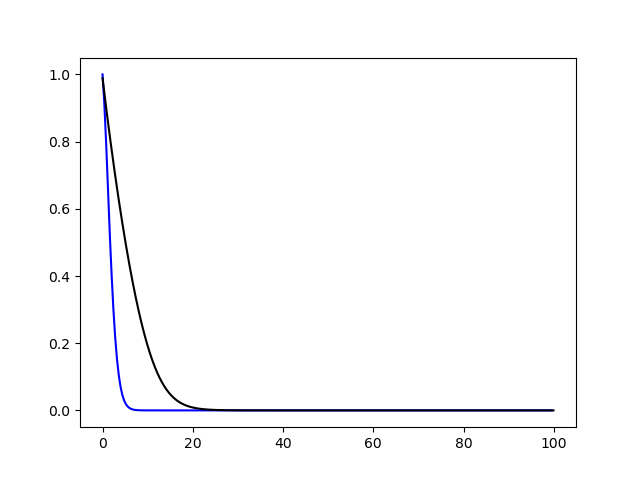}
\end{subfigure}\hfil
\begin{subfigure}{0.3\textwidth}
\includegraphics[width=0.9\linewidth, height=3cm]{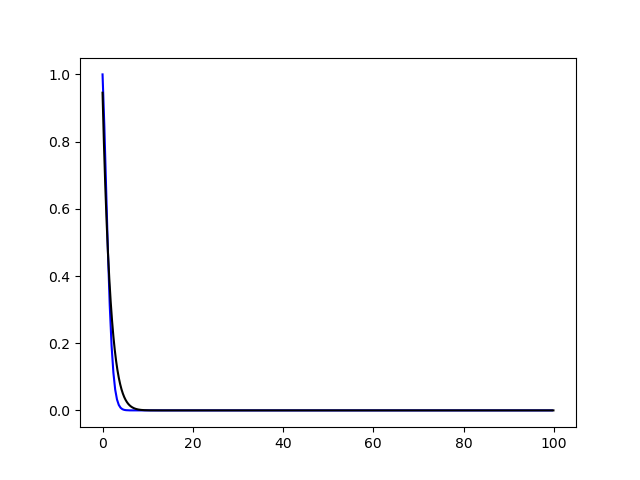}
\end{subfigure}\hfil
\begin{subfigure}{0.3\textwidth}
\includegraphics[width=0.9\linewidth, height=3cm]{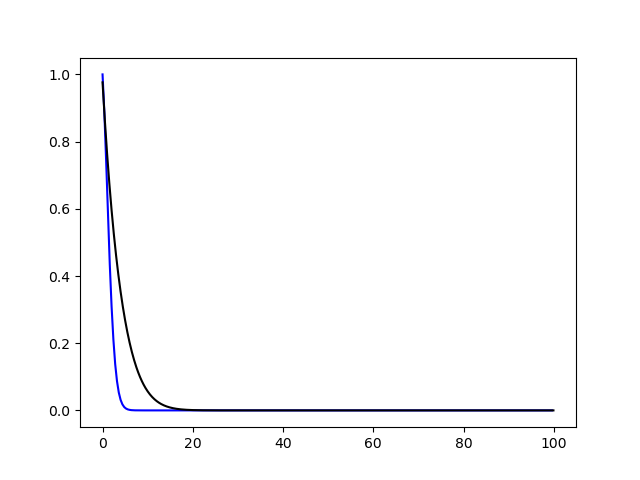} 
\end{subfigure}\hfil
\begin{subfigure}{0.3\textwidth}
\includegraphics[width=0.9\linewidth, height=3cm]{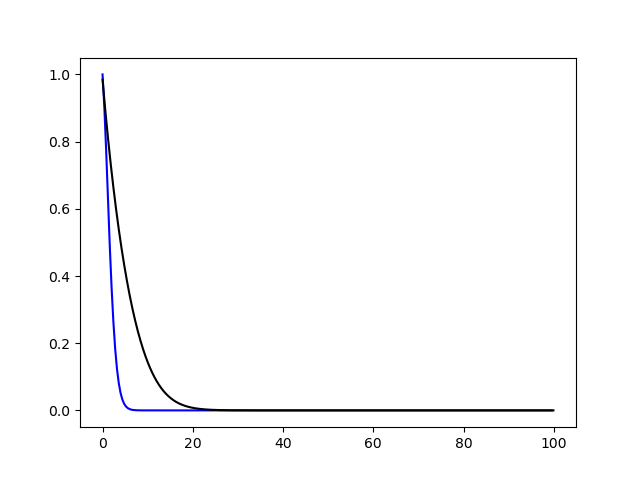} 
\end{subfigure}\hfil
\begin{subfigure}{0.3\textwidth}
\includegraphics[width=0.9\linewidth, height=3cm]{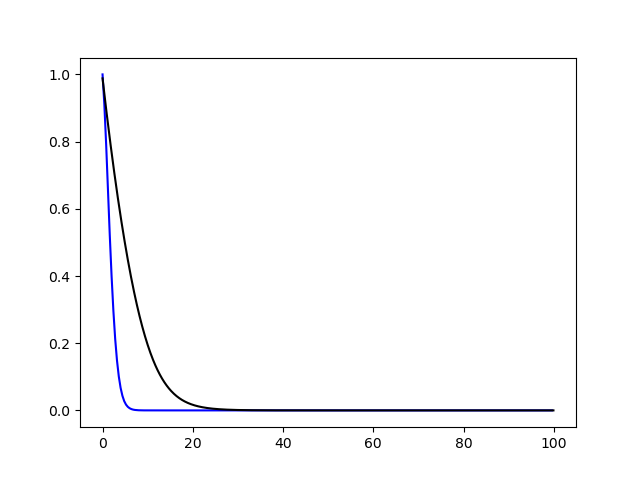} 
\end{subfigure}\hfil
\begin{subfigure}{0.3\textwidth}
\includegraphics[width=0.9\linewidth, height=3cm]{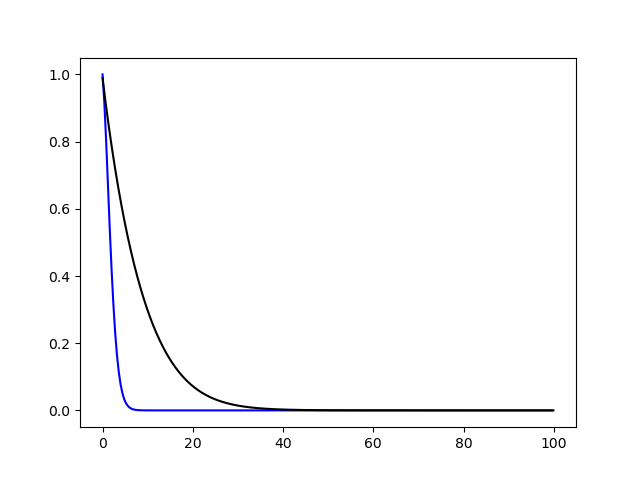} 
\end{subfigure}\hfil
\begin{subfigure}{0.3\textwidth}
\includegraphics[width=0.9\linewidth, height=3cm]{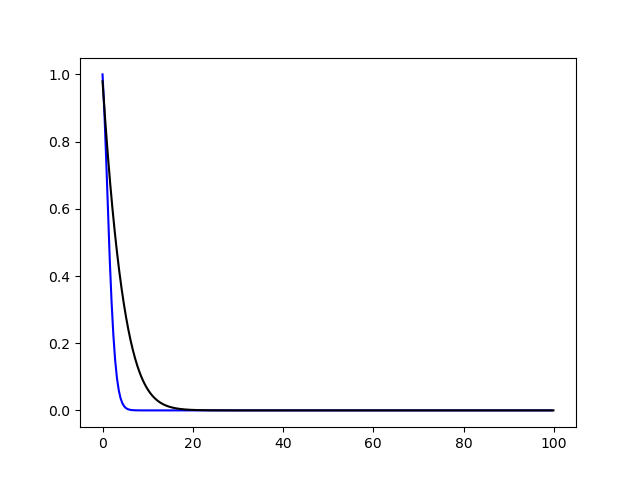} 
\end{subfigure}\hfil
\caption{Conditional survival curves for 9 different sets of covariates estimated by the Cox model with sample size $n = 2000$.}
\label{fig:arbitrary_cox}
\end{figure}

Figure~\ref{fig:arbitrary} shows the results of fitting the DeepONet with FNN branch net and Figure~\ref{fig:arbitrary_cnn} shows the results with CNN based branch net. We see that the sample average of corresponding estimated curves (yellow dash line) using either method well overlaps with the ground truth (black solid line), but using CNN structure gives slightly narrower confidence bands. Figure~\ref{fig:arbitrary_cox} shows the biased results obtained from the misspecified Cox model, where only the sample averages are presented. The large bias of the Cox model is expected because the instantaneous covariate effect assumption is violated in this simulation setup.

\begin{figure}[htb!]
\centering
\begin{subfigure}{0.3\textwidth}
\includegraphics[width=0.9\linewidth, height=3cm]{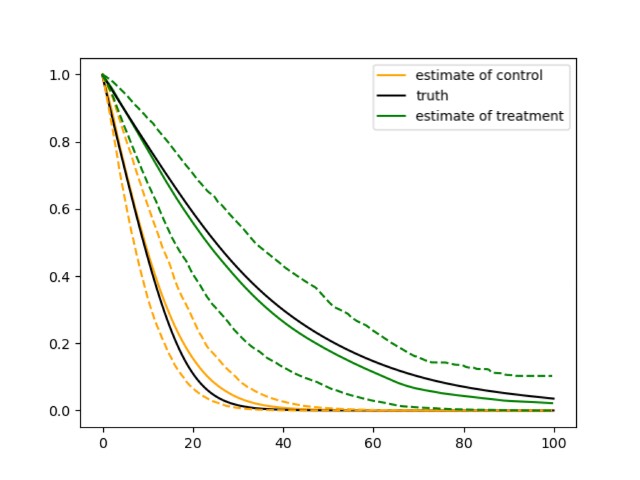} 
\end{subfigure}\hfil
\begin{subfigure}{0.3\textwidth}
\includegraphics[width=0.9\linewidth, height=3cm]{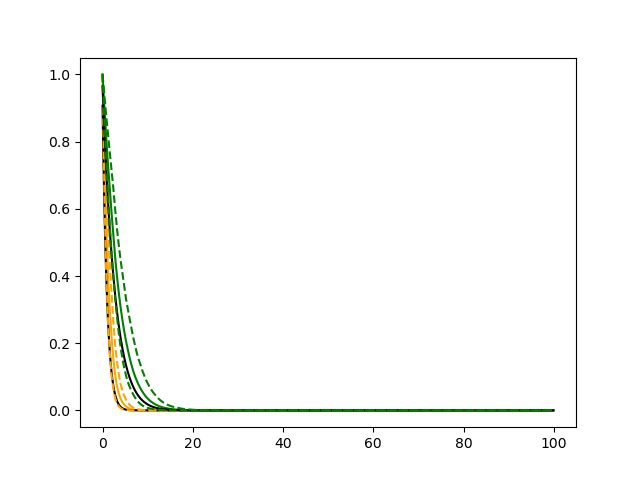}
\end{subfigure}\hfil
\begin{subfigure}{0.3\textwidth}
\includegraphics[width=0.9\linewidth, height=3cm]{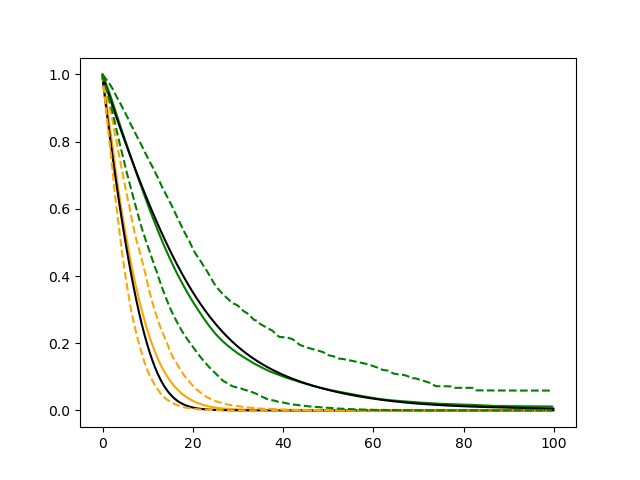}
\end{subfigure}\hfil
\begin{subfigure}{0.3\textwidth}
\includegraphics[width=0.9\linewidth, height=3cm]{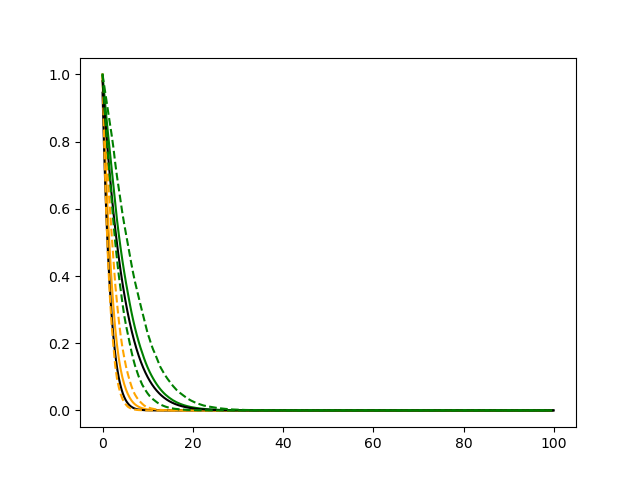}
\end{subfigure}\hfil
\begin{subfigure}{0.3\textwidth}
\includegraphics[width=0.9\linewidth, height=3cm]{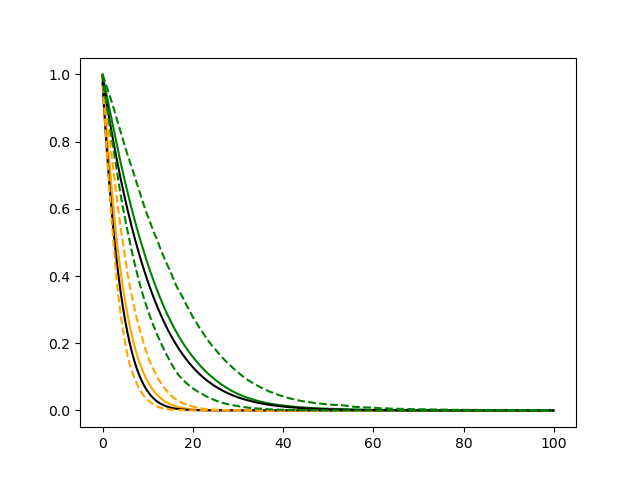} 
\end{subfigure}\hfil
\begin{subfigure}{0.3\textwidth}
\includegraphics[width=0.9\linewidth, height=3cm]{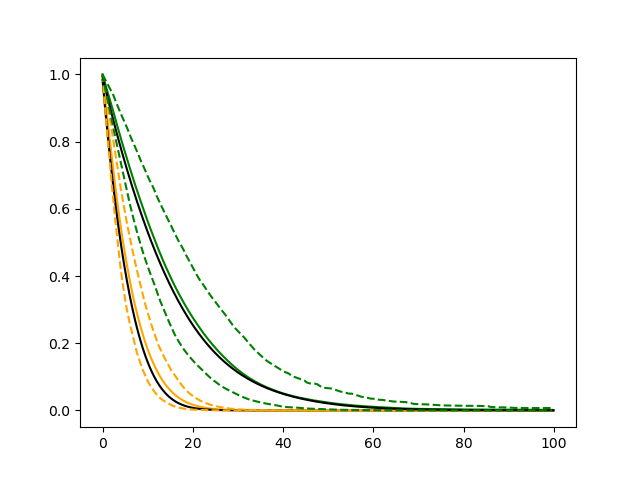} 
\end{subfigure}\hfil
\begin{subfigure}{0.3\textwidth}
\includegraphics[width=0.9\linewidth, height=3cm]{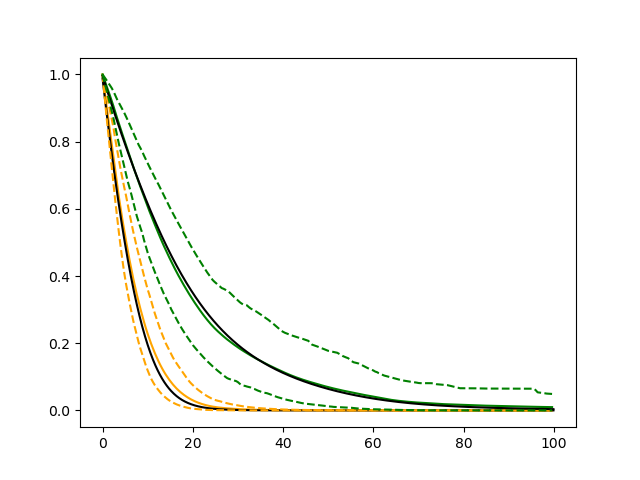} 
\end{subfigure}\hfil
\begin{subfigure}{0.3\textwidth}
\includegraphics[width=0.9\linewidth, height=3cm]{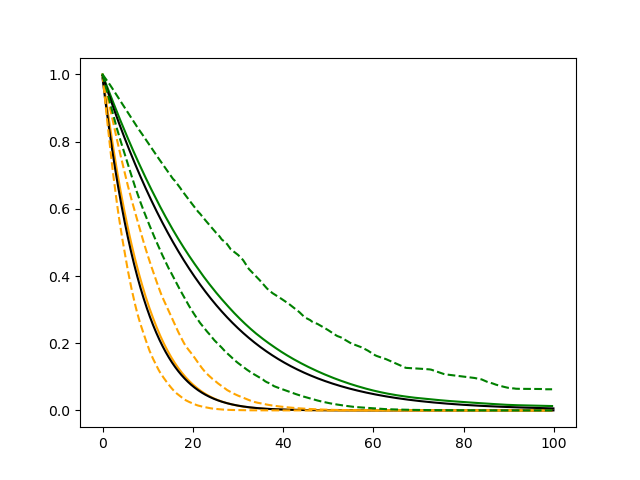} 
\end{subfigure}\hfil
\begin{subfigure}{0.3\textwidth}
\includegraphics[width=0.9\linewidth, height=3cm]{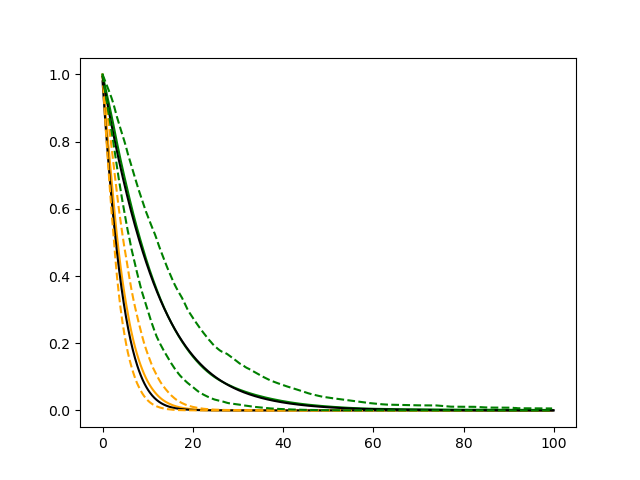} 
\end{subfigure}\hfil
\caption{Conditional survival curves corresponding to $z_i =0$ and $z_i=1$, respectively, for 9 different sets of covariates $x_i(t)$ and $w_i$, mimicking 9 individuals.}
\label{fig:trt_ctr}
\end{figure}

For each of these 9 sets of covariates, we can obtain two estimated survival curves corresponding to $z_i = 0$ (control group) and $z_i = 1$ (treatment group), respectively, mimicking a setup of estimating the individual level treatment effects nonparametrically given $x_i(t)$ and $w_i$. 
This also provides a framework of adjusting for potential confounders nonparametrically in causal inference without imposing any model assumptions for the survival data. Figure~\ref{fig:trt_ctr} illustrates such estimates, where the solid green curve represents the sample average of the estimated survival curves when $z_i=1$ and the solid yellow curve represents the sample average of the estimated survival curves when $z_i=0$. The black solid curves are the corresponding true survival curves. We can see that the treatment effect varies among different individuals, reflecting the nature of our simulation setup.

\bigskip
\begin{center}
[Figures 2-5 are around here.]
\end{center}
\bigskip

\section{ADNI Data Analysis}

We analyze data from the Alzheimer's Disease Neuroimaging Initiative (ADNI) (https://adni.loni.usc.edu) to illustrate our method. ADNI data are collected to study the progression of Alzheimer's disease in the human brain. Specifically, we use ANDIMERGE, which is a merged version of ADNI data, with a data cutoff on January 4, 2025. For our analysis, we select participants who were diagnosed with mild cognitive impairment (MCI) at baseline and who had at least one follow-up visit. The event of interest is defined as the progression to dementia, and the corresponding outcome is progression-free survival. In this context, we consider Mini-Mental State Examination (MMSE) score and hippocampal volume to be external time-varying covariates, and include baseline age as a time-invariant covariate. After excluding individuals with missing baseline MMSE or hippocampal volume, the final dataset consists of observations from 849 individuals. Time-varying covariate values are taken on a six-month interval grid, and missing interim values are imputed using forward-filling. The maximum follow-up duration is 198 months. 

Given the sparsity of time points in this real-world dataset, we implement the FNN architecture with a single hidden layer as the branch net in our DeepONet framework. The number of nodes in the dense layer is 128, and the learning rate is 0.0001. Instead of using mini-batches, we train the model on the whole data. For comparison, we also fit a Cox proportional hazards model assuming instantaneous time-varying covariate effects. We evaluate the model performance using the integrated Brier score (IBS) \citep{Graf1999AssessmentAC} based on a 5-fold cross-validation. Our method achieves a lower cross-validated mean IBS of 0.20 compared to 0.28 by the Cox model. The reduction in IBS highlights the effectiveness of our model in capturing complex patterns in the data that the Cox model may not fully account for.

\section{Discussion}

It is worth noting that our proposed neural network-based full likelihood approach is fully nonparametric with the weakest assumptions, thus providing the maximum robustness against model assumptions. Any method with efficiency gain under stronger conditions will suffer from the risk of model misspecification.

A possible limitation of the proposed method is that, when incorporating covariate history into the neural network, we construct an expanded dataset where the covariate history is paired with every time point. 
This approach, while straightforward, increases the memory requirement due to the large volume of data involved. Improving the efficiency of data handling and storage is crucial, which is of great interest for future research.


\bibliographystyle{unsrtnat}
\bibliography{main}

\pagebreak

\begin{table}[htb!]
    \centering
    \begin{tabular}{ccccc}
    \hline
    $i$ & time point & covariates\\
    \hline
    1  & $t_1$ & $x_1(t_0), 0, \dots, 0 $\\
    1  & $t_2$ & $x_1(t_0), x_1(t_1)$, 0, \dots, 0\\
    \vdots & \vdots & \vdots\\
    1  & $t_m$ & $x_1(t_0), x_1(t_1), ..., x_1(t_{m-1})$ \\
    2  & $t_1$ & $x_2(t_0), 0, \dots, 0  $\\
    2  & $t_2$ & $x_2(t_0), x_2(t_1), 0, \dots, 0 $\\
    \vdots & \vdots & \vdots\\
    2  & $t_m$ & $x_2(t_0), x_2(t_1), ..., x_2(t_{m-1})$ \\
    \vdots &  \vdots & \vdots \\
    \hline
    \end{tabular}
    \caption{The data structure for time-varying covariates.}
    \label{tab:data_structure_arbitrary}
\end{table}

\end{document}